\documentclass[aps,prl,showpacs,twocolumn]{revtex4}
\usepackage{bm,color,amsmath,amssymb,mathrsfs,latexsym,graphicx,psfrag}

\newcommand{\bk}{{\mathbf k}}
\newcommand{\bq}{{\mathbf q}}
\newcommand{\be}{\begin{equation}}
\newcommand{\ee}{\end{equation}}

\def\be{\begin{equation}}
\def\ee{\end{equation}}
\def\bea{\begin{eqnarray}}
\def\eea{\end{eqnarray}}

\def\C60{A$_x$C$_{60}$}

\def\HgCu3{HgCa$_2$Cu$_3$O$_{8+y}$}
\def\HgCu4{HgBa$_2$Ca$_3$Cu$_4$O$_{10+y}$}
\def\TlCu{Tl$_2$Ba$_2$CuO$_{6+\delta}$}
\def\TlCu3{Tl$_2$Ba$_2$Ca$_2$Cu$_3$O$_{10+y}$}
\def\TlCu4{Tl$_2$Ba$_2$Ca$_3$Cu$_4$O$_{12+y}$}

\def\BiCu3{Bi$_2$Sr$_2$Ca$_{2}$Cu$_3$O$_y$}

\def\8LSCO{La$_{1.88}$Sr$_{.12}$CuO$_4$}
\def\110LNSCO{La$_{1.5}$Nd$_{0.4}$Sr$_{0.1}$CuO$_{4}$}
\def\stage4LCO{La$_{2}$CuO$_{4+\delta}$}
\def\Y248{YBa$_2$Cu$_4$O$_8$}

\def\NbSe2{NbSe$_2$}
\def\TaSe2{TaSe$_2$}
\def\TiSe2{TiSe$_2$}

\begin{document}

\title{Multi-Weyl Topological Semimetals Stabilized by Point Group Symmetry}

\author{Chen Fang$^{1}$, Matthew J. Gilbert$^{2,3}$, Xi Dai$^{4}$,  B. Andrei Bernevig$^{1}$}
\affiliation{$^1$Department of Physics, Princeton University, Princeton NJ 08544}
\affiliation{$^2$Department of Electrical and Computer Engineering, University of Illinois,  Urbana IL 61801, USA}
\affiliation{$^3$Micro and Nanotechnology Laboratory, University of Illinois, 208 N. Wright St, Urbana IL 61801, USA}
\affiliation{$^4$Institute of Physics, Chinese Academy of Sciences, Beijing 100080, China}
\date{\today}

\begin{abstract}

We perform a complete classification of two-band $\bk\cdot\mathbf{p}$ theories at band crossing points in 3D semimetals with $n$-fold rotation symmetry and broken time-reversal symmetry. Using this classification, we show the existence of new 3D topological semimetals characterized by $C_{4,6}$-protected double-Weyl nodes with quadratic in-plane (along $k_{x,y}$) dispersion or $C_6$-protected triple-Weyl nodes with cubic in-plane dispersion. We apply this theory to the 3D ferromagnet HgCr$_2$Se$_4$ and confirm it is a double-Weyl metal protected by $C_4$ symmetry. Furthermore, if the direction of the ferromagnetism is shifted away from the [001]- to the [111]-axis, the double-Weyl node splits into four single Weyl nodes, as dictated by the point group $S_6$ of that phase. Finally, we discuss experimentally relevant effects including splitting of multi-Weyl nodes by applying $C_n$ breaking strain and the surface Fermi arcs in these new semimetals.
\end{abstract}

\maketitle

The discovery and classification of topological states of matter beyond those found in time reversal invariant (TRI) materials\cite{Kane:2005sf,Kane2005,Bernevig:2006kx,Fu:2007fk,Haldane1988,Moore:2007uq,bernevig2006a} is vital to understanding the full spectrum of possible topological phases. In principle, it is possible to find many more topological systems each containing distinct physical observables characterized by the symmetries they preserve. Non-trivial topology in systems possessing crystallographic point group symmetries (PGS) is of particular interest\cite{Sun:2009,Turner:2010,Hughes:2011,Fu:2011} because PGS universally exist in solids. And a classification based on PGS may encompass time-reversal breaking magnetically ordered materials that are useful for potential applications.

Most recently, 3D topological \emph{semimetals} have been proposed to exist in a variety of materials such as cold atom systems\cite{Sun:2010,Jiang:2012}, multilayer topological insulator systems\cite{Halasz:2011,Burkov:2011A} and pyrochlore iridates\cite{Phosur:2010,xwan:2011}.  Unlike in a topological insulator, the band structure of a topological semimetal exhibits bulk gapless points in the Brillouin zone (BZ). In close proximity of the gapless points, the effective Hamiltonian is that of a 3D Weyl fermion therefore these bulk gapless points are referred to as Weyl nodes\cite{xwan:2011}. In 3D translationally invariant systems, Weyl nodes are protected from opening a gap against infinitesimal transformations of the Hamiltonian; these points act as monopoles (vortices) of 3D Berry curvature, as any closed 2D surface surrounding them exhibits a unit Chern number, and they can be gapped only by annihilation with other Weyl points of opposite monopole charge. They may exist both in the presence of TRI, as in the metallic state between a TRI trivial and a nontrivial insulator, and in its absence (pyrochlore irridates). Without the presence of any other symmetries, Weyl nodes are the generic topological band-crossings in 3D semimetals.

In this letter, we show theoretically that a series of new 3D topological semimetals may exist when PGS are present in a material. Physically, this is because PGS can bring together two or multiple Weyl nodes with nonzero net monopole charge together onto a high-symmetry point, resulting in double- (quadratic in two directions) or multiple-crossings in the BZ. These crossings, which we hereafter denote as ``double-Weyl node'' or ``triple-Weyl node', are protected from splitting into Weyl nodes by the PGS. We investigate these protected crossings starting from a classification of all two-band $\bk\cdot\mathbf{p}$ theories at high-symmetry points in 3D crystals with $C_n$ point group. By applying the classification in 2D, we prove in 2D insulators a relation between the Chern number (mod $n$) and all $C_m$ eigenvalues of occupied bands at $C_m$ invariant $\bk$-points in BZ, where $m$ divides $n$. By applying the classification to 3D semimetals, we determine the type of each band-crossing point by knowing the symmetry representations of the conduction and valence bands on high-symmetry lines. As a result, $C_4$ or $C_6$ symmetry can protect double-Weyl nodes while only $C_6$ symmetry can protect triple-Weyl nodes and there cannot be any higher order crossings protected by $n$-fold rotation symmetries. We use this to analyze the recently proposed 3D semimetal HgCr$_2$Se$_4$\cite{Gangxu:2011} in the ferromagnetic (FM) phase and find it a $C_4$ protected double-Weyl metal with two double-Weyl nodes along $\Gamma{Z}$.

Consider a 3D crystal that is invariant under an $n$-fold rotation about $z$-axis, where by lattice restriction $n=2,3,4,6$, and containing no other symmetries. $C_n$-invariance also implies $C_m$ invariance for any $m$ divides of $n$. In a tight-binding model with translational symmetry, $C_m$-invariance gives: \bea\hat{C}_m\hat{H}(\bk)\hat{C}^{-1}_m=\hat{H}(R_m\bk)\label{eq:general},\eea where $\hat{C}_m$ is the $m$-fold rotation operator and $R_m$ is the $3\times3$ rotation matrix defining the 3D $m$-fold rotation. For any $C_m$, we define a $C_m$ invariant line on which $R_m\mathbf{k}=\mathbf{k}$ is satisfied at every $\mathbf{k}$. Besides the rotation axis, there exist additional rotation invariant lines due to periodicity of BZ. On any $C_m$-invariant line, Eq.(\ref{eq:general}) implies $[\hat{C}_m,\hat{H}(\mathbf{k})]=0$, such that all bands on the line may be labeled by the corresponding eigenvalues of $\hat{C}_m$. If the conduction and valence bands are very close in energy  at $\mathbf{K}$ on a $C_m$-invariant line, we can approximate the effective Hamiltonian around that point by a $2\times2$ matrix:
\bea H_{eff}(\mathbf{K}+\bq)=f(\bq)\sigma_++f^*(\bq)\sigma_-+g(\bq)\sigma_z,\label{eq:Heff_general}\eea where $\mathbf{q}$ is assumed to be small and in-plane ($q_z=0$). The above Hamiltonian is written in the basis where $(1,0)^T$ represents the Bloch wavefunction for the conduction band at $\mathbf{K}$ and $(0,1)^T$ represents the valence band. In the above equation $f$ is a complex function, and $g$ a real function; and $\sigma_\pm=\sigma_x\pm i\sigma_y$. In this basis, the matrix representation of $\hat{C}_m$, $\mathcal{C}_m$, is a diagonal matrix with $(\mathcal{C}_{m})_{11}=u_c$, $(\mathcal{C}_m)_{22}=u_v$. From Eq.(\ref{eq:general},\ref{eq:Heff_general}) and the explicit form of $\mathcal{C}_m$, we can obtain all symmetry constraints on the functional forms of $f$ and $g$, however, the full proof is relegated to the Supplementary Material. The constraints and the effective theories to the lowest order of $\bq$ are summarized in Table \ref{tab:Heff}. In particular, we note that the constraint on $g$ always takes the form $g(q^+,q_-)=g(q_+e^{i2\pi/m},q_-e^{-i2\pi/m})$. This allows a non-zero $\bq$- independent term, $g(0)=m(K_z)$. Therefore, if we know the critical wavevector at which $m(K_c)=0$, then we have a 3D-node and the functional forms of $f$ determine the nature of that node.

Among various classes of nodes shown in the Table \ref{tab:Heff}, several cases deserve special attention as they describe Weyl nodes featuring quadratic and cubic dispersion in $q_{x,y}$ and carrying $\pm2$ and $\pm3$ monopole charge, respectively. Consider $u_c/u_v=-1$ at $K_z=K_c$ on a $C_4$-invariant line, then the effective Hamiltonian, to lowest order, reads, \bea H_{eff}(\bq)=(aq_+^2+bq_-^2)\sigma_++h.c,\label{eq:doubleWeyl}\eea where $a,b$ are arbitrary complex numbers. Due to the absence of linear terms in Eq. (\ref{eq:doubleWeyl}), this Hamiltonian describes a double-Weyl node. On a $C_6$ invariant line, if $u_c/u_v=-e^{\pm i2\pi/3}$, we again have a double-Weyl node at $K_z=K_c$. If, however, $u_c/u_v=-1$, at $K_z=K_c$ on a $C_6$ invariant line, then the effective Hamiltonian reads, \bea H_{eff}(\bq)=(aq_+^3+bq_-^3)\sigma_++h.c.\eea This describes a triple-Weyl node as both linear and quadratic terms are absent. We find that no higher order Weyl node beyond triple is protected in 3D. These high-symmetry nodes in a 3D semimetal with $C_n$ symmetry may be identified by examining the band structure along all $C_4$ and $C_6$ invariant lines and evaluating $(u_c,u_v)$ at each crossing point (if any).

\begin{table*}
\begin{tabular}{|c|c|c|c|c|}
\hline
$m$ & $u_c/u_v$ & Constraints on $f$ & $H_{eff}$ & Q\\
\hline
$2$ & $-1$ & $f(-q_+,-q_-)=-f(q_+,q_-)$ & $m\sigma_z+(aq_++bq_-)\sigma_++h.c.$ & $sign(|a|-|b|)$\\
\hline
$3$ & $e^{\pm{i}2\pi/3}$ & $f(q_+e^{i2\pi/3},q_-e^{i2\pi/3})=e^{\pm i2\pi/3}f(q_+,q_-)$ & $m\sigma_z+ak_\pm\sigma_++h.c.$ & $\pm 1$\\
\hline
$4$ & $\pm{i}$ & $f(iq_+,-iq_-)=\pm if(q_+,q_-)$ & $m\sigma_z+aq_\pm\sigma_++h.c.$ & $\pm 1$\\
    & $-1$ & $f(iq_+,-iq_-)=-f(q_+,q_-)$ & $m\sigma_z+(aq_+^2+bq_-^2)\sigma_++h.c.$ & $2sign(|a|-|b|)$\\
\hline
$6$ & $e^{\pm i\pi/3}$ & $f(q_+e^{i\pi/3},q_-e^{i\pi/3})=e^{\pm i\pi/3}f(q_+,q_-)$ & $m\sigma_z+aq_+\sigma_++h.c.$ & $\pm1$\\
    & $e^{\pm i2\pi/3}$ & $f(q_+e^{i\pi/3},q_-e^{i\pi/3})=e^{\pm i2\pi/3}f(q_+,q_-)$ & $m\sigma_z+aq_\pm^2\sigma_++h.c.$ & $\pm2$\\
    & $-1$ & $f(q_+e^{i\pi/3},q_-e^{i\pi/3})=-f(q_+,q_-)$ & $m\sigma_z+(aq_+^3+bq_-^3)\sigma_++h.c.$ & $3sign(|a|-|b|)$\\
\hline
\end{tabular}
\caption{All two-band $\bk\cdot\mathbf{p}$ theories on $C_m$ invariant lines for $m=2,3,4,6$ for all possible combinations of $(u_c,u_v)$. $m$ is a real parameter and $a,b$ are complex parameters which depend on $K_z$. Here we provide the general constraints on $f(q_+,q_-)$ where $q_\pm=q_x\pm iq_y$. It should be noted that the constraints for $g$ take a general form: $g(q_+e^{i2\pi/m},q_-e^{i2\pi/m})=g(q_+,q_-)$ and is therefore suppressed.}
\label{tab:Heff}
\end{table*}

Aside from classifying multi-Weyl nodes in 3D semimetals, the results presented in Table \ref{tab:Heff} have important implications in 2D Chern insulators. Any $k_z$-slice ($k_z\neq{k_c}$) of a 3D system is a 2D insulator with $k_z$ as a parameter. One may perform a continuous interpolation between different 2D insulators and the interpolation can be mapped onto a 3D BZ using `$k_z$' as the interpolation parameter. Between a 2D Chern insulator and a trivial insulator, the interpolation must have band crossings, which map to nodes in the fictitious 3D BZ. Via Gauss's Law, the Chern number of the Chern insulator equals the net monopole charge between the `planes' defining the trivial and Chern insulators. To be more concrete, consider a $C_4$ invariant Chern insulator on a square lattice with a band crossing at either $\Gamma$ or $M$ with $C_4$ eigenvalues $(u_c,u_v)$ ($(u_v,u_c)$) right before (after) the crossing. If $u_c/u_v=i$, then, from Table \ref{tab:Heff}, we obtain a charge of $+1$. This means that the Chern number increases by one as the $C_4$ eigenvalue of the valence band changes by a factor of $i$. If $u_c/u_v=-1$, then the charge is $\pm2$, and the Chern number changes by $\pm2$ as the $C_4$ eigenvalue changes by a factor of $-1$. On the other hand, if the band crossing is at $C_2$ invariant point $X$, then the charge is always $\pm1$, and, as there are two $X$'s in the BZ, the total Chern number change must be $\pm2$. We may concisely summarize all Chern number changing scenarios for $C_4$ in one compact formula as
\bea\label{eq:ChernC4}\exp(i2\pi C/4)=\prod_{n\in{occ}}(-1)^F\xi_n(\Gamma)\xi_n(M)\zeta_n(X),\eea where $\xi_n$ and $\zeta_n$ are the $C_4$ and $C_2$ eigenvalues on the $n$th band, respectively while $F=0$ ($F=1$) denotes spinless (spinful) fermions. The derivation may be repeated for $C_{n=2,3,6}$ insulators to obtain
\bea\label{eq:ChernC3}\exp(i2\pi C/3)&=&\prod_{n\in{occ}}(-1)^F\theta_n(\Gamma)\theta_n(K)\theta_n(K'),\\
\label{eq:ChernC6}\exp(i2\pi C/6)&=&\prod_{n\in{occ}}(-1)^F\eta_n(\Gamma)\theta_n(K)\zeta_n(M),\eea where $\theta_n$ are the $C_3$ eigenvalues and $\eta_n$ are the $C_6$ eigenvalues on the $n$th band. (The definitions of high-symmetry points: $\Gamma=(0,0)$, $X=(\pi,0)$, $M=(\pi,\pi)$ in $C_4$ invariant systems, $K=(0,4\pi/3\sqrt{3})$, $K'=(0,-4\pi/3\sqrt{3})$ and $M=(2\pi/3,0)$ in $C_{3,6}$ invariant systems.)

\begin{figure}[!hbt]
\includegraphics[width=8cm]{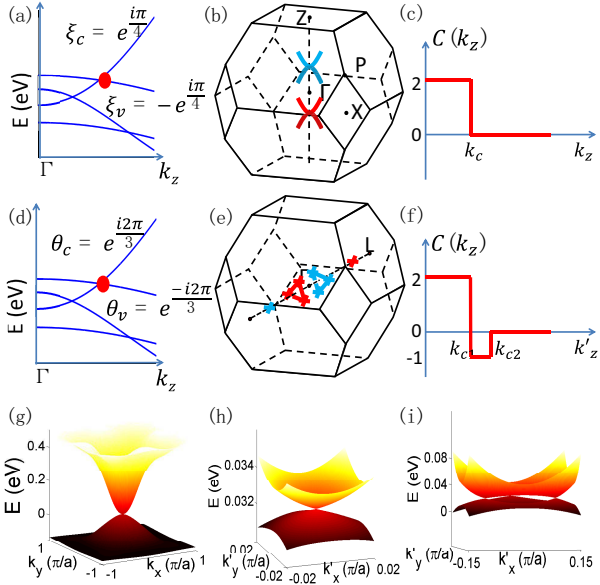}
\caption{(a)/(d) Band structure along $\Gamma{Z}$/$\Gamma{L}$ in 3D FM HgCr$_2$Se$_4$ with magnetization $\mathbf{M}$ parallel to $[001]$/$[111]$ directions, obtained from an eight-band tight-binding model fitted from LDA results. The $C_4$/$C_3$ eigenvalues of the conduction and the valence bands are also shown. (b)/(e) The schematic showing the two double-Weyl points/four Weyl points in 3D BZ with $\mathbf{M}$ parallel to $[001]$/$[111]$. Red/blue means nodes with positive/negative monopole charge(s). (c)/(f) The schematic of the Chern number as a function of $k_z$/$k'_z$ along $\Gamma{Z}$/$\Gamma{L}$ with $\mathbf{M}$ parallel to $[001]$/$[111]$. (g) The conduction and the valence band dispersion at $k_z=k_c=0.43$ with $\mathbf{M}$ parallel to $[001]$. (h) and (i) The conduction and the valence band dispersion at $k'_z=k_{c1}=0.43$ and $k'_z=k_{c2}=0.15$ with $\mathbf{M}$ parallel to $[111]$, respectively.}
\label{fig:theory}
\end{figure}

We apply the above general theory to a 3D FM HgCr$_2$Se$_4$. HgCr$_2$Se$_4$ possesses point group $O_h$ in the paramagnetic phase which breaks down to $C_{4h}$ in the FM phase. This indicates that the system is invariant under a four-fold rotation about $z$-axis, and a mirror reflection about the $xy$-plane. Bandstructure calculations show no band crossings along $XP$, while along $\Gamma{Z}$, there is a single band-crossing at $k_z=k_c=0.43a^{-1}$. In Fig.\ref{fig:theory}(a), the band structure is plotted with the respective $C_4$ eigenvalues along $\Gamma{Z}$ and the red point marks the crossing between conduction and valence bands with $u_c/u_v = -1$. The crossing point is thus a double-Weyl node with either monopole charge of $\pm 2$. This node \emph{cannot} split into two single Weyl nodes either along $z$-axis or in $xy$-plane as long as the $C_4$ symmetry is unbroken. The mirror symmetry about $xy$-plane ensures the existence of another double-Weyl node with opposite monopole charge of $\mp 2$. In Fig.\ref{fig:theory}(b), we plot a schematic of the two double-Weyl nodes in the BZ. The quadratic dispersion along $k_x$ and $k_y$ is confirmed by the calculated band dispersion shown in Fig.\ref{fig:theory}(g). That such double-Weyl node corresponds to a jump of $\pm2$ in Chern number at $k_z=k_c$ is confirmed in Fig.\ref{fig:theory}(c), where we calculate the Chern number on a nearest neighbor hopping eight-band tight-binding model on the FCC lattice for HgCr$_2$Se$_4$\cite{Gangxu:2011}.

Additionally, the existence of an $xy$-mirror plane implies, for a 3D system, that $H(k_x,k_y,-k_z)=\mathcal{M}_{xy}H(k_x,k_y,k_z)\mathcal{M}^{-1}_{xy}$. Hence, at every $(k_x,k_y)$, the Hamiltonian $H(k_x,k_y,k_z)$ possesses 1D inversion symmetry ($k_z\rightarrow{k_z}$) which quantizes the electric polarization to be $P_z(k_x,k_y)=0$ or $P_z(k_x,k_y)=1/2$ depending on the formula $(-1)^{2P_z}=\prod_{i\in{occ.}}\zeta_i(0)\zeta_i(\pi)$\cite{Hughes:2011}. Change of $P_z$ from $0$ to $1/2$ or vice versa implies crossing a bulk node. In HgCr$_2$Se$_4$ we find that at a point very close to $k_x=k_y=0$, $P_z(0\pm,0)=1/2$ (at $k_x=k_y=0$, the system is gapless at $k_z=\pm{k_c}$), while at a corner of the BZ, $P_z(\pi,0)=0$. Therefore, along any path connecting $(0\pm,0)$ and $(\pi,0)$, there must be a node. As the path is arbitrary, the result is a line node in the BZ that can only appear on the $k_z=0$ or $k_z=\pi$ plane and is \emph{protected} by $C_{4h}$ symmetry.

Microscopically, band crossings in HgCr$_2$Se$_4$ are due to the $s$-$p$ orbital inversion in the FM phase\cite{Gangxu:2011}, but we will show such inversion alone does not guarantee the existence of double-Weyl nodes. In Fig.\ref{fig:theory}(d), we plot the band structure along $\Gamma{L}$ with ferromagnetism oriented along the $[111]$-axis (which we now denote as the new $z'$ direction, in a system $(x',y',z')$). Fig.\ref{fig:theory}(d) is quite similar to Fig.\ref{fig:theory}(a), however, the PGS has changed from $C_{4h}$ to the six-fold rotation-reflection about $z'$, or $S_6=C_6*M_{x'y'}$. This symmetry is different from $C_6$ and our general theory cannot be directly applied. But noting that $S_6^2=C_3$, we can calculate the $C_3$ eigenvalues of the conduction and valence bands along $\Gamma{L}$, and find $(u_c,u_v)=(e^{-i2\pi/3},e^{i2\pi/3})$ (marked in Fig.\ref{fig:theory}(d)). This crossing point has charge $+1$, and is, therefore, \emph{not} a double-Weyl node. Furthermore, on the planes $k'_z=0$ and $k'_z=\sqrt{3}\pi/2a$, which are invariant under $M_{x'y'}$ due to the periodicity of BZ, $C_6$ symmetry is recovered. Using Eq. (\ref{eq:ChernC6}) to evaluate their Chern numbers, we find that $C(k_z'=0)=6n+2$ and $C(k_z'=\sqrt{3}\pi/2a)=6n$, which implies a net charge of $-2$ between the two planes. The Weyl node along $\Gamma{L}$ at $k_{c1}$ contributes $+1$, so there must be three other Weyl nodes at $k_{c2}$ related to each other by $C_3$, each having charge $-1$. The correct configuration of the bulk nodes is shown in Fig.\ref{fig:theory}(e). In Figs.\ref{fig:theory}(h) and (i), we plot the linear dispersion around the Weyl point at $k_{c1}$ and the three Weyl points at $k_{c2}$. In going from $\Gamma$ to $L$, the Chern number takes two jumps of $-3$ and $+1$ (see Fig.\ref{fig:theory}(f)). Physically, the splitting of a double-Weyl node in this case is due to spin-orbital interaction which is only compatible with $S_6$ but not $C_4$ invariance.

\begin{figure}[!hbt]
\includegraphics[width=8cm]{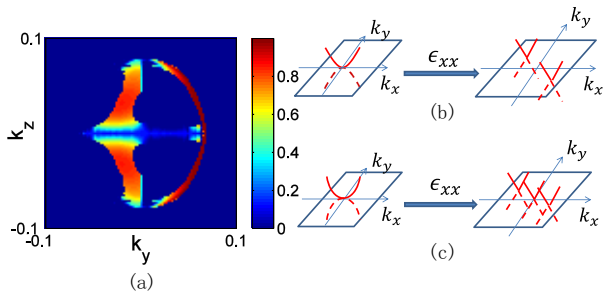}
\caption{(a) Momentum resolved spectral weight ($\propto\textrm{Im}[G(k_y,k_z,E)]$) calculated on the surface of FM HgCr$_2$Se$_4$ after the system is cut along the $k_x$-direction. The energy is varied as a function of $k_z$ to ensure that it is inside the bulk gap. The intensity is normalized such that unity means the state is completely localized on the surface. (b) and (c) Schematics showing that a double- and a triple- Weyl node breaks down to two and three Weyl nodes under an applied strain $\epsilon_{xx}$, respectively.}\label{fig:exp}
\end{figure}

In Fig.\ref{fig:exp}(a) we plot the double-Fermi arcs on the $(100)$-surface of FM HgCr$_2$Se$_4$. Like TRI topological insulators, 3D topological semimetals have surface states. Yet the surface states that cross the Fermi energy do not form a closed loop as in other 2D systems, but appear as Fermi arcs. Specifically, in a Weyl-semimetal, if one measures the electron spectral weight on the $(mnl)$-plane at the Fermi energy, one will see a number of arcs, each of which connects projected images of the two bulk nodes of opposite charge on $(mnl)$-plane. When we extend the picture to double/triple-Weyl semimetals, there must be two/three Fermi arcs connecting the two projected images of the two bulk nodes of opposite charge, on the $(mnl)$-plane. The exception to this occurs when the $[mnl]$-direction is the rotation axis, when the arcs collapse into a single Fermi point.

Another potentially fruitful way to experimentally probe the properties of multi-Weyl semimetals is to examine the quantum phase transitions induced by an applied strain. Since the multi-Weyl nodes studied here are protected by $C_n$-invariance, an applied strain that breaks such symmetry can split a multi-Weyl node into several single Weyl nodes. To see how such splitting may happen, consider adding a $C_4$ breaking term in the effective Hamiltonian around a double-Weyl node Eq.(\ref{eq:doubleWeyl}):
\bea\Delta{H}=(\epsilon_{xx}-\epsilon_{yy})\sigma_x.\eea One can check that this term changes sign under $\mathcal{C}_4\propto\sigma_z$ but is invariant under $C_2=C_4^2$, justifying its coupling to an anisotropic strain $\epsilon_{xx}-\epsilon_{yy}$. The dispersion of the adapted Hamiltonian can be easily solved and two nodes emerge at \bea \bq_{\pm,c}=(\pm\delta{q}\cos\theta,\pm\delta{q}\sin\theta,K_c)\eea, where $\theta$ satisfies $|a|\sin(\theta_a+2\theta)=|b|\sin(2\theta-\theta_b)$ and $\delta{q}=\sqrt{\epsilon(|a|\cos(\theta_a+2\theta)+|b|\cos(2\theta-\theta_b))/2}$. Straightforward calculation shows that both nodes are Weyl nodes having \emph{equal} monopole charge $sign(|a|-|b|)$. In Fig.\ref{fig:exp}(b,c), we show, schematically, how a double-/triple-Weyl node breaks into two/three Weyl nodes under an anisotropic strain in the $xy$-plane. This transition is in general marked by the change of density of states ($\rho(E)$) near the node, which in principle affects bulk transport properties. For example, $\rho(E)\propto|E-E_c|$ in double-Weyl semimetals, while $\rho(E)\propto(E-E_c)^2$ after the splitting by strain, where $E_c$ is the energy at the node.

\begin{acknowledgments}
The authors thank T. L. Hughes, A. Alexandradinata, Zhong Fang and D. Haldane for fruitful discussions. CF acknowledges travel supported by the ONR under grant N0014-11-1-0728 and salary support from ONR - N00014-11-1-0635. MJG acknowledges support from the AFOSR under grant FA9550-10-1-0459 and the ONR under grant N0014-11-1-0728 and a gift the Intel Corporation. BAB was supported by NSF CAREER DMR- 095242, ONR - N00014-11-1-0635, Darpa - N66001-11- 1-4110 and David and Lucile Packard Foundation. Xi Dai is supported by NSF China and the 973 program of China (No. 2011CBA00108).
\end{acknowledgments}
%

\end{document}


\onecolumngrid
\section{Explicit proof of constraints and effective theories presented in Table I}

In this appendix, we give a general derivation the constraints and effective theories shown in Table I. On a $C_m$ invariant line in a $C_n$ invariant system ($m$ being a factor of $n$), each band corresponds to a 1D representation of cyclic group $C_m$ (spinless), or double cyclic group $C^D_m$ (spinful). In either case, an eigenvalue of $C_m$ is in the form:\bea\alpha_p=\exp(i2\pi p/m+iF\pi/m),\eea where $p=0,1,...m-1$. Now suppose the $C_m$ eigenvalues of the conduction and valence bands are\bea u_c=\alpha_p,\\
\nonumber u_v=\alpha_q.\eea We have if $p=q$, then the two bands will \emph{not} cross on this $C_m$-invariant line because of the presence of a symmetry allowed a constant off-diagonal term $\delta|\psi_u\rangle\langle\psi_v|+h.c.$ which can always open a gap. If $p\neq q$, then the matrix representation of $C_m$, $\mathcal{C}_m$ is given by\bea\mathcal{C}_m=\exp(i\pi\frac{F+p+q}{m})\exp(i\pi\frac{p-q}{m}\sigma_z).\eea The transform of $H_{eff}(\bq)$ under $C_m$ is given by\bea\label{eq:temp1}\mathcal{C}_m{}H_{eff}(\bq)\mathcal{C}^{-1}_m&=&g(\bq)\sigma_z+f(\bq)\exp(i\pi\frac{p-q}{m}\sigma_z)\sigma_+\exp(-i\pi\frac{p-q}{m}\sigma_z)+h.c.\\
\nonumber&=&g(\bq)\sigma_z+f(\bq)e^{-i2\pi(p-q)/m}\sigma_++h.c.\eea
In the basis of $q_\pm$, the $R_m\bq$ is given by\bea R_m(q_+,q_-)=(q_+e^{i2\pi/m},q_-e^{-i2\pi/m}).\label{eq:temp2}\eea Substituting Eq.(\ref{eq:temp1},\ref{eq:temp2}) into Eq.(1) in main text, we obtain\bea e^{-i2\pi(p-q)/m}f(q_+,q_-)=f(q_+e^{i2\pi/m},q_-e^{-i2\pi/m})\label{eq:temp3},\\
\nonumber g(q_+,q_-)=g(q_+e^{i2\pi/m},q_-e^{-i2\pi/m}).\eea This is the general constraint on $f,g$ by $C_m$ symmetry for a given pair of $(u_c,u_v)$. From Eq.(\ref{eq:temp3}) we learn that the forms of $f$ and $g$ only depend on $p-q$, or $u_c/u_v$. These results comprise the central column of Table I.

To generate the last column from the general constraints Eq.(\ref{eq:temp3}), we start from an expansion of \bea f(q_+,q_-)=\sum_{n_1n_2}A_{n_1n_2}q_+^{n_1}q_-^{n_2},\eea where $A_{n_1n_2}$ is an arbitrary complex coefficient. Eq.(\ref{eq:temp3}) gives $A_{n_1n_2}=0$ if $n_2-n_1\neq p-q\;\textrm{mod}\;m$. Then we pick up $(n_1,n_2)$ with smallest $n_1+n_2$ and nonzero $A_{n_1n_2}$ to obtain the last column of Table I.

Physically, one can easily understand Table I as the consequence of the `semi-conservation' of total angular momentum $J_z$. Here `semi-conservation' means that $J_z$ is only conserved up to a multiple of $m$. This is because in a lattice the continuous rotation symmetry downgrades to discrete rotation symmetry of order $m$. If $u_c/u_v=e^{i2\pi(p-q)/m}$, if means the conduction and valence bands differ by $\delta J_z=p-q$. Therefore, the off-diagonal term of the effective Hamiltonian, $|\psi_c\rangle\langle\psi_v|$, must couple a $k_-^{p-q}$ or $k_+^{m-p+q}$ to conserve the total angular momentum up to a multiple of $m$.